# Harnessing Network Science for Urban Resilience: The CASA Model's Approach to Social and Environmental Challenges

Miguel Fuentes [1,2,3,4], Juan Pablo Cárdenas [2,5], Gastón Olivares [2,5], Eric Rasmussen [6], Carolina Urbina [7], Soledad Salazar [2], Gerardo Vidal [2,8]

*1. Santa Fe Institute, Santa Fe, NM, USA.*
*2. Complex Society Lab, Viña Del Mar, Chile.*
*3. IIF-SADAF, Buenos Aires, Argentina.*
*4. Instituto de Sistemas Complejos de Valparaíso, Valparaíso, Chile.*
*5. Net-Works, Viña Del Mar, Chile.*
*6. Infinitum Humanitarian Systems, Seattle, WA, USA.*
*7. Escuela de Psicología, Pontificia Universidad Católica de Valparaíso, Viña Del Mar, Chile.*
*8. Pontificia Universidad Católica de Valparaíso, Valparaíso, Chile.*

Corresponding author: Juan Pablo Cárdenas

## Abstract

Resilience in social systems is crucial for mitigating the impacts of crises, such as climate change, which poses an existential threat to communities globally. As disasters become more frequent and severe, enhancing community resilience has become imperative. This study introduces a cutting-edge framework, quantitative network-based modeling called Complex Analysis for Socio-environmental Adaptation (CASA) to evaluate and strengthen social resilience. CASA transforms resilience models' linear and static structure into a complex network that integrates complexity and systems thinking, utilizing global scientific knowledge and complex network methodologies. The resulting resilience framework features rich interdependencies, and subsequent dimensionality reduction produces robust resilience indicators. This innovative application of network sciences is then demonstrated by quantitatively assessing what are known as "Sacrifice Zones," socio-environmentally sensitive areas. Results unveil the potential of this novel application of complex network methodologies as tools for systemic diagnostics, identifying vulnerabilities, and guiding policies and practices to enhance climate resilience and adaptation. The CASA framework represents a pioneering tool for assessing territorial resilience, leveraging network science applications, big data analytics, and artificial intelligence. CASA serves as a systemic diagnostic tool for urban resilience and a guide for policymakers, urban planners, and other professionals to promote sustainable, healthy cities in an era of climate change.

## 1.- Introduction

Resilience, essential in the current climate of heightened vulnerability, refers to the ability to bounce back from adversity, a concept extensively discussed by Fraccascia, Giannoccaro, and Albino (2018). This quality is not innate but developed through a system's interactions and experiences. Systems with a more extensive history of such interactions will likely demonstrate superior adaptability and resilience, emerging more robust from challenges. In Complex Systems Theory, systems that assimilate and apply information gleaned from environmental interactions are recognized as Complex Adaptive Systems (CAS), a concept delineated by John H. Holland (1992) and Murray Gell-Mann (2018), among others.

The concept of resilience extends beyond its basic definition to encompass varying scales within systems, ranging from individual groups to organizations, communities, and even entire territories, regions, or states (Lengnick-Hall et al., 2011; Maguire & Cartwright, 2008). Resilient individuals are characterized by their ability to make informed decisions, often grounded in quantitative analysis, and to derive meaning and personal growth from challenging experiences. These individual traits can significantly influence and enhance the resilience of larger systems, particularly when decision-makers are involved. According to Shi et al. (2021), resilient organizations exhibit similarities to resilient individuals, including a defined sense of purpose, a commitment to learning from past errors, and cultivating a culture prioritizing open communication and collaboration. A resilience approach identifies resources and capabilities (Maguire & Cartwright, 2008). Such an environment is critical for fostering resilience at a community level, facilitating the seamless and transparent exchange of information essential for effectively tapping adaptive capacity.

The interaction between individual and community resilience characteristics can be addressed from a deep look at the conditions and challenges of specific territories. A territory can be defined as a set of historically structured social interactions, geographically located and constantly evolving. More precisely, integrating a model that identifies the capacities installed in a social system implies the construction of territorial and community resilience (Williner & Tognoli, 2023).

In social systems, resilience is closely linked with robustness and antifragility. Robustness allows these systems to withstand stresses without disruption, while antifragility enables them to grow stronger from such challenges (Johnson & Gheorghem, 2013). The concept of a Complex Adaptive System (CAS) provides a valuable lens for studying resilience in social environments, given their self-organizing and adaptive nature in response to external influences. Social systems, exhibiting properties like diversity and modularity typical of CAS, adapt and evolve through environmental interactions (Batty, 2009; Shi et al., 2021). Viewing social systems as CAS, resilience is defined by their ability to absorb and recover from pressures while maintaining their structural integrity. This involves flexibility, redundancy, and establishing diverse connections within the system. Understanding the behavior of these systems necessitates grasping the complexity arising from component interactions that can lead to unpredictable, nonlinear dynamics.

From a CAS perspective, various evaluation methods can be employed to assess urban systems' underlying complexity and dynamics, particularly their resilience (Bozza, Asprone & Manfredi, 2015; Rus, Kilar & Koren, 2018). Recognizing their intricate feedback loops and nonlinear dynamics is crucial for realistic and effective urban planning and management strategies. It is equally important to highlight that social resilience models are fundamental to international relations and science diplomacy for various reasons, all of which are related to the ability of societies to confront and adapt to global challenges. This adaptability is an essential prerequisite for international stability and cooperation. From a scientific perspective, these models provide a framework for understanding how communities can survive, adapt, and thrive in the face of adversities, which is crucial for formulating effective diplomatic policies and strategies (Folke, 2005; Ruffini & Ruffini, 2017).

Before community resilience can be enhanced, its current state must be assessed, and the *Analysis of Resiliency for Integrated System Effectiveness* (ARISE) framework was a useful starting point. Our choice was based on the comprehensiveness of the model (to be discussed in detail later), which demonstrates the potential to include various elements present in contemporary society. ARISE is a product of Dr. John Hummel's lab at Argonne National Laboratory in the United States. Hummel (2019) defines resilience as the capacity of an entity—be it an asset, organization, community, or region—to anticipate, resist, absorb, respond to, adapt to, and recover from disruptions, whether natural or human-induced. The ARISE framework collects and orders data to help judge that capacity.

In this paper, we employ the basic framework of ARISE to construct a new organizational structure for its elements, sectors, and pillars, considering global scientific knowledge. The Complex Analysis for Socio-environmental Adaptation (CASA) model introduced in this paper is the outcome of this process, encapsulating the fundamental interactions within the social system that determine its resilience. The CASA model implies the possibility of knowing the capacity of a social system, its territory, and its institutions to face the adversities caused by disasters and the effects of climate change or environmental degradation. In this way, it would seek to improve community resilience to reorganize itself so that its functions, structures, and identities are strengthened.

This is our core working hypothesis. When enhanced by specific data on the capabilities installed in each territory, the CASA topology can become a powerful tool for analyzing the capacities within a social system.

The paper's structure is as follows: The next section introduces the original ARISE model and outlines the methodology for transforming it into the CASA model. The third and fourth sections provide an analysis of the proposed model's structure and the outcomes of applying this model to a territory facing significant environmental pressures. The final section outlines our conclusions and the research opportunities afforded by this methodological approach.

## 2.- Methodology

Developed a decade ago as a straightforward but extensive spreadsheet, the ARISE framework provides a tool for evaluating resilience levels by collecting extensive public (non-PII) data sources and manipulating the resulting values to form indices. It was initially used by the *Center of Excellence in Disaster Management and Humanitarian Assistance* (COE-DMHA) within Tripler Army Medical Center in Hawaii to assess the resilience of Pacific Rim countries. The values derived from an ARISE analysis provide a resilience snapshot for a point in time and can suggest strategies for strengthening the assessed communities.

The ARISE framework recognizes that promoting regional resilience involves considering regional challenges in an increasingly complex and interwoven landscape. That landscape involves geography, economies, natural resources, demographics, and culture. ARISE recognizes that understanding such diverse data as logistics flow, emergency response systems, and social media is essential when developing resilience strategies in communities and their territories. It proved an excellent starting point for introducing increasing complexity in social and infrastructural relationships over time and recognizing the interplay of forces shaping societies under pressure.

This current study started with ARISE and then explored the interplay between resilience and systemic complexity, specifically in the climate adaptation processes within a community. It begins with analyzing the ARISE model's strengths and weaknesses, followed by a method for transforming ARISE into a next-generation framework to support climate adaptation resilience: *Complex Analysis for Socio-environmental Adaptation* (CASA). Then, CASA is applied in the context of climate change's impact on populations, concluding with its implications for territorial planning and the promotion of adaptive, flexible strategies informed by complex systems methodologies and artificial intelligence tools.

The original model ARISE is structured around five primary pillars, as illustrated in Fig. 1. This approach originally mirrored the methodology employed by the Center of Excellence for Disaster Management and Humanitarian Assistance (COE DMHA) to evaluate healthcare systems in developing nations. Under each pillar, resilience is categorize into distinct sectors, which further break down into specific components, offering a detailed framework for assessing and enhancing resilience at various levels.

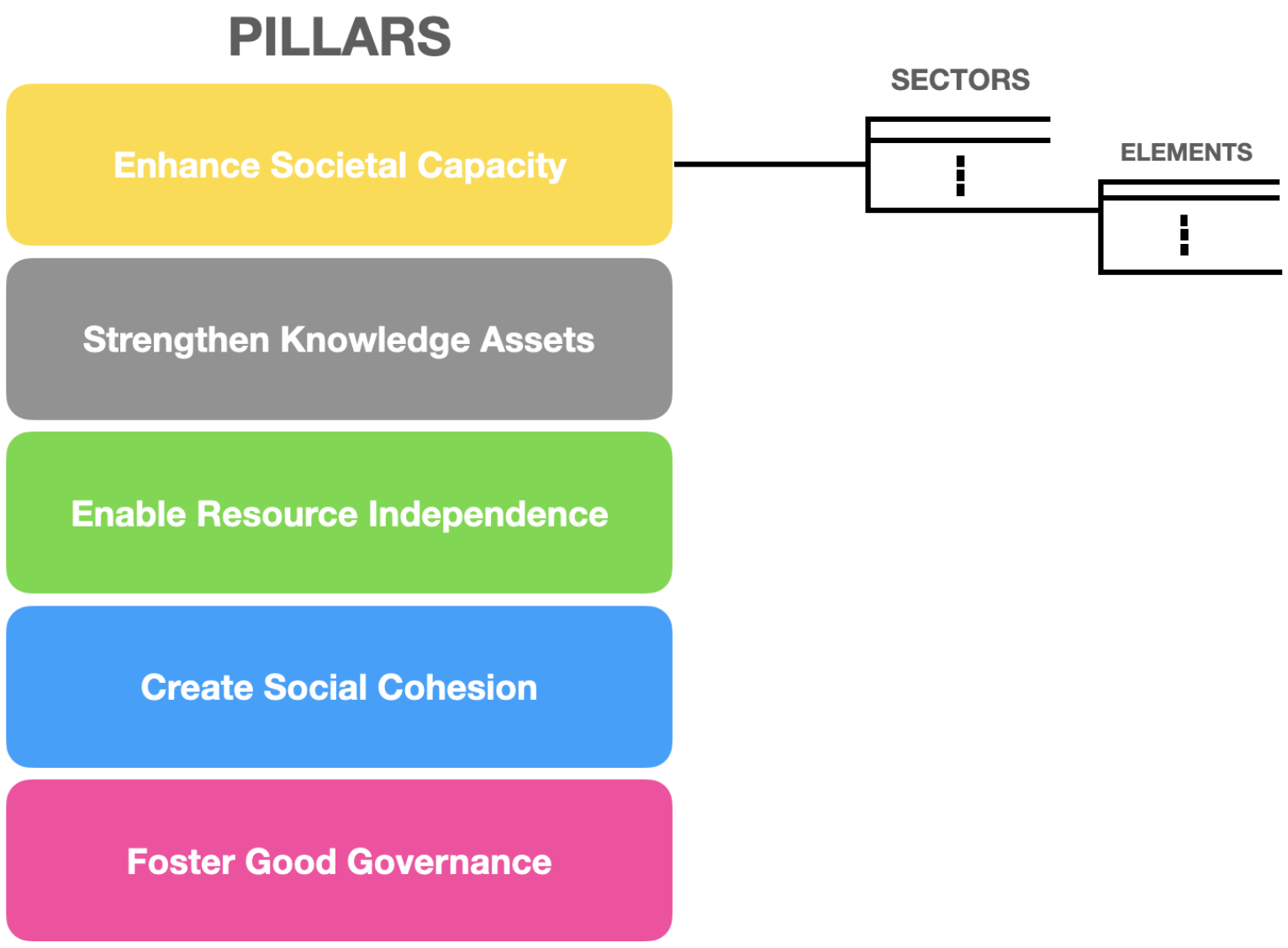

**Figure 1.** The fundamental structure of the ARISE model.

One or more indicators have been established within each element. In cases where multiple indicators exist within a resilience element, their values can be weighted to provide an assessment that considers their relative importance within the area to be studied.

The ARISE framework adopts a fairly rigid structural tree model (see, for example, the breakdown of the "Enhance Societal Capacity" pillar in Fig. 2), where each element influences only its respective sector and pillar. This fundamentally linear design is unsuited to the dynamic complexity of climate change adaptation and ignores interdependencies among the framework's elements.

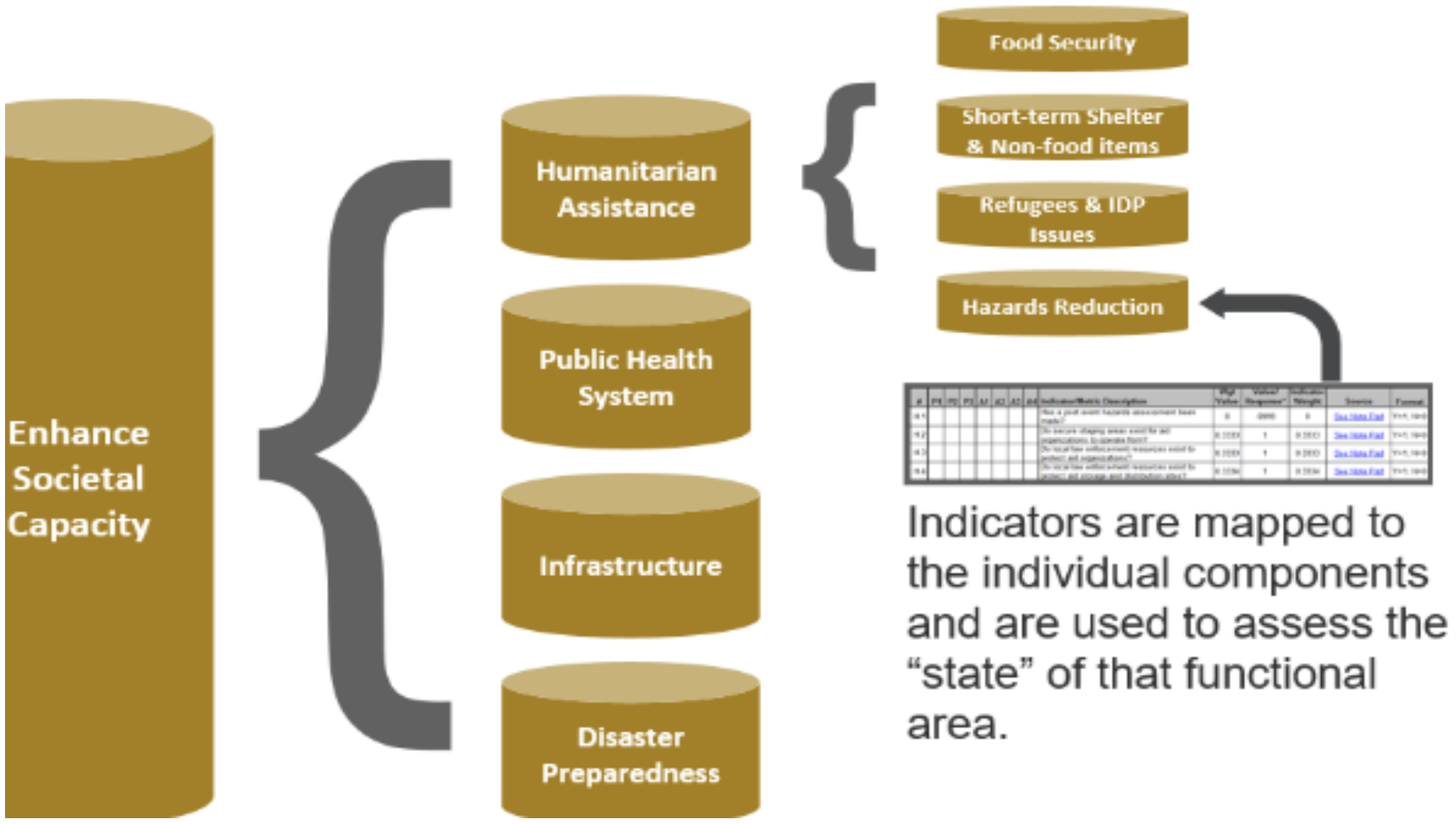

**Figure 2.** Breakdown of the "Enhance Societal Capacity" pillar.

The ARISE framework is underpinned by data from many primary sources. These primary sources predominantly include reliable and consistent information from esteemed national and international entities, such as the World Health Organization and the World Bank. Such sources offer a blend of qualitative and quantitative data, which undergoes thorough evaluation before its release.

Next-tier sourcing involves research-grade data tailored to specific themes from organizations recognized for their expertise. An illustrative example would be the Fragile States Index, an annual report published by the Fund for Peace. Such research indices typically use proprietary analytical methodologies and are often geared towards highlighting the challenges faced by a nation or region.

The lowest tier of data sources encompasses news outlets, government publications, and social media platforms. Although these channels furnish timely information, they are prone to biased inaccuracies and may exhibit subjective reporting. The compilation of all tiers of data feeding into the ARISE framework is ultimately converted into indicators that contribute to formulating a territorial resilience index, reflecting a region's capacity to withstand and adapt to various challenges.

### 2.1 The CASA Model

As mentioned above, the ARISE framework exhibits a tree-like structure consisting of three distinct levels: pillars, sectors, and elements. While seemingly straightforward, the autonomy of components within each tier leads to unrecognized consequences that are not visualized. This detachment among elements primarily challenges forecasting future scenarios involving interactions, cascade effects, and feedback loops. This could also hinder understanding the ramifications of alterations in one component on others within the same level, potentially instigating further changes across various levels. Secondly, it is vital to formulate indicators that capture disparate elements not inherently connected in the ARISE framework, including those from other sectors or pillars. The ARISE model's relatively simplistic structural design limits the systemic methodology required by a contemporary grasp of complexity - an aspect our study intends to address. Thus, it is imperative to acknowledge this potential complexity and transition to a more network-centric analytical framework, building upon the ARISE model's core elements and sources to obtain the required data.

Therefore, for constructing the CASA Model, we advocate for a novel model grounded in complex systems theory, which perceives the social system of a territory as an adaptive entity exhibiting evolutionary characteristics. Methodologically, we propose a model integrating interactions among the foundational components of any tree-like structure model, thereby establishing a complex network of elements, sectors, and pillars. System complexity typically relates to its size, configuration, and the number of parameters or links. Increasing complexity can be beneficial in certain scenarios, and enhancing the model with a selection of the methods mentioned below appears advantageous:

- **Increase the number of layers:** Adding more layers to a network increases its depth, allowing it to learn hierarchical representations of data. Deeper networks often capture more complex relationships and improve performance on indicators requiring high abstraction. However, since the ARISE model already comprises three layers (pillars, sectors, and elements), further complexity through this method is deemed unnecessary for this study.

- **Increase the number of units per layer**: Expanding the number of units in each layer enhances the network's capacity to model complex relationships within the data. However, significantly larger networks may become computationally burdensome, harder to comprehend for decision-makers, and more susceptible to overfitting. Since the ARISE framework already addresses a large percentage of the known dimensions around the resilience of a territory, in this study, we decided not to modify this dimension.

- **Increase the number of connections or parameters:** Introducing more connections between units can also elevate the network's complexity. Additional connections enable the representation of new and more sophisticated indicators. This is the strategy we have chosen for this work, allowing for increased interactions at the same level as the units, sectors, and pillars within the ARISE model.

While increasing network complexity can enhance the new model's performance, it should be approached judiciously. As noted earlier, overly complex networks may result in computational cost, difficulty comprehending results for decision-makers, and a higher risk of overfitting scenarios. Therefore, balancing complexity and generalization ability is crucial, considering specific requirements and the available human and computational resources. In the following section, we will describe the procedure used to generate a more realistic and complex model for territorial resilience, considering the interactions between elements within any tree-like structure model.

### 2.1.1 A Scientifically Informed Connectivity

As described above, the ARISE model is an evaluative tool for assessing territorial resilience, employing structured inquiries. These inquiries are methodically arranged into Elements, further classified into Sectors, and ultimately aggregated into Pillars. This work aims to enhance the ARISE model's complexity by incorporating an advanced approach to synthesizing contemporary global scientific knowledge. For this, we have carefully selected 61 specific elements from the ARISE model to establish a comprehensive systemic framework. We conducted exhaustive literature searches in scientific databases to elucidate the intricate interconnections among these elements and construct an interconnected network. Our methodology involved the application of various network analysis techniques previously utilized in studies such as the Knowledge Mapping of Chile (Cárdenas et al., 2015), the Mapping and Analysis of Social and Scientific Networks (Cárdenas et al., 2010; Fuentes et al., 2023; Cárdenas et al., 2014), and a research tool developed by our team (DataCiencia, 2023).

A significant challenge we faced was the potential ambiguity of element names in the ARISE model during global database searches, notably in the Web of Science (WoS), an important source in this study. To overcome this ambiguity, we compiled a keyword dictionary for each element derived from the foundational questions of the ARISE model. These queries were rigorously examined and refined to more concise and recognizable keywords, effectively capturing the essence of each element.

The next step was an exhaustive review of documents within the WoS database involving keyword combinations related to pairs of elements. This approach enabled us to define the element network as a comprehensive, undirected network, represented as G(E, L), where E signifies the array of distinct elements, and L indicates their co-occurrence in the Title, Abstract, WoS Keywords, and Author's Keywords of scientific articles globally from 2008 to 2023. The result was an analysis of more than 70 million documents.

### 2.1.2 Network Density Reduction

The initial network, designated as G(E, L), displayed a remarkably high density, incorporating numerous links, even in those with minimal weights. This level of complexity presented significant challenges for network analysis and interpretation. We applied the Maximum Spanning Tree (MST) algorithm to the network to address this. The MST algorithm is specifically formulated to construct a tree within a weighted network. Its goal is to connect all network elements using the heaviest possible links. This approach aims to maintain the overall structure of the network while focusing on the most substantial and influential relationships, thereby eliminating less significant and relevant connections.

While the implementation of the MST effectively reduced the network's density, it also necessitated the removal of certain links that might hold considerable importance. Acknowledging this limitation, we deemed establishing additional sector-based networks to preserve and examine these significant relationships crucial.

Subsequently, a decision was made to create a sector network, denoted as H(S,L), derived from the original element network G(E,L). This network aimed to represent relationships at a higher level of aggregation. In this context, "S" signifies the set of sectors, and "L" represents their co-occurrence relationships. To facilitate this transformation, a dictionary was used to map elements to their respective sectors, reducing the network's scope from 61 elements to 18 sectors.

Notably, the MST was not applied to H(S,L) to retain the integrity of sectorial relationships. A secondary network, labeled H', was then generated by applying the MST algorithm to the sector network H(S,L), as illustrated in Fig.3. This procedure produced a minimal tree connecting all sectors, emphasizing relationships with the highest weights.

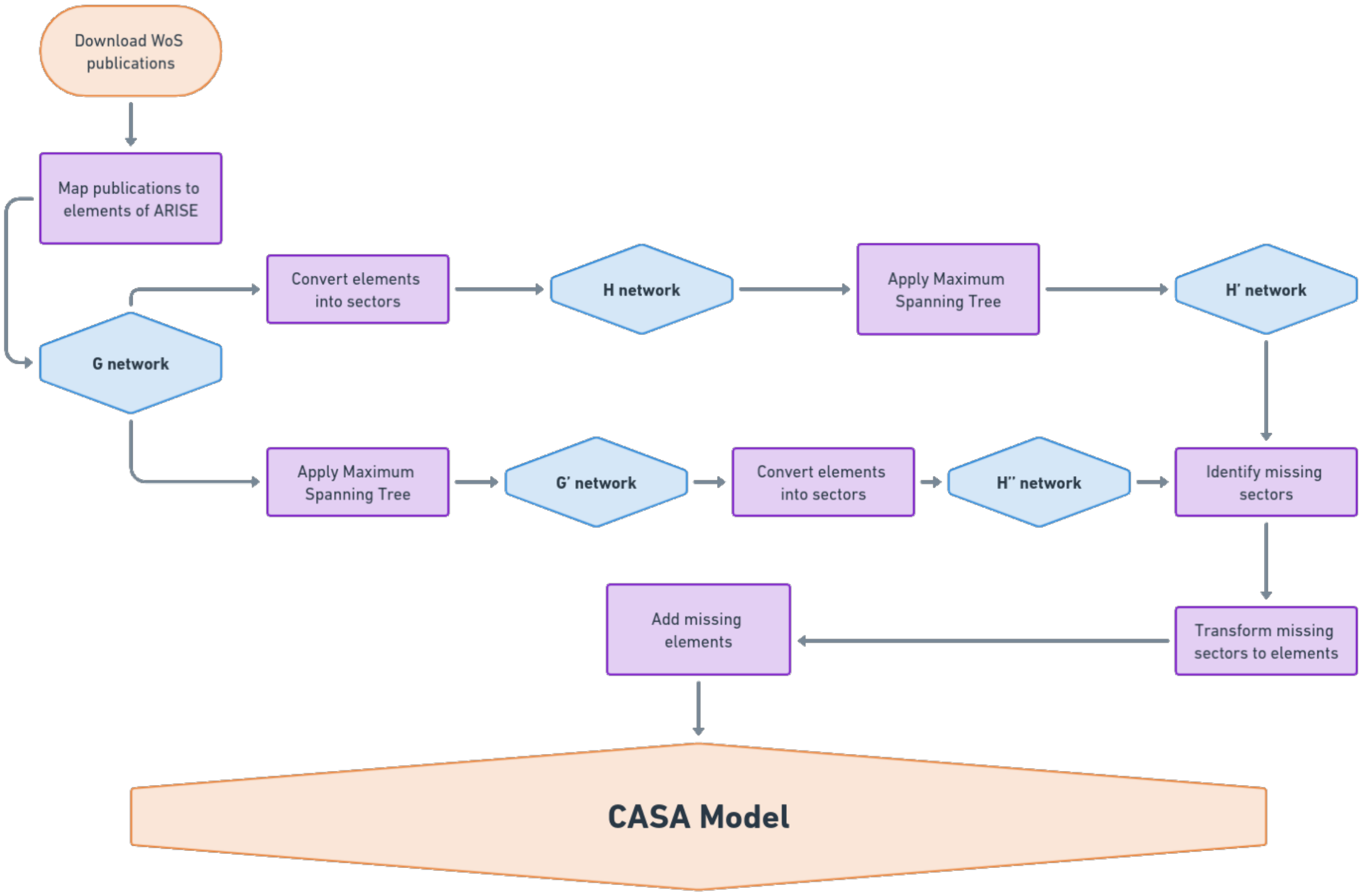

**Figure 3.** Diagram of the methodology used to develop the CASA model. Refer to the accompanying text for a detailed description of how the *Network Density Reduction* process, among other strategies, efficiently minimized the number of interactions employed.

The MST algorithm was also utilized on the G network to derive a streamlined element network, G'. From G', a third sector network, labeled H'', was constructed using the same transformation dictionary.

We identified pairs of sectors absent in both H' and H'' networks. These pairs, marked as "s," signify relationships excluded during the MST process but potentially highly important. An inverse dictionary transformation was executed to identify key elements within the "s" pairs to address this. These pivotal elements were manually integrated into the G' network, enriching the depiction of significant relationships in the final network structure.

This process concluded the sector network generation phase, allowing a more nuanced and accurate analysis of sector and element relationships within the maturing CASA methodology framework (Fig. 4).

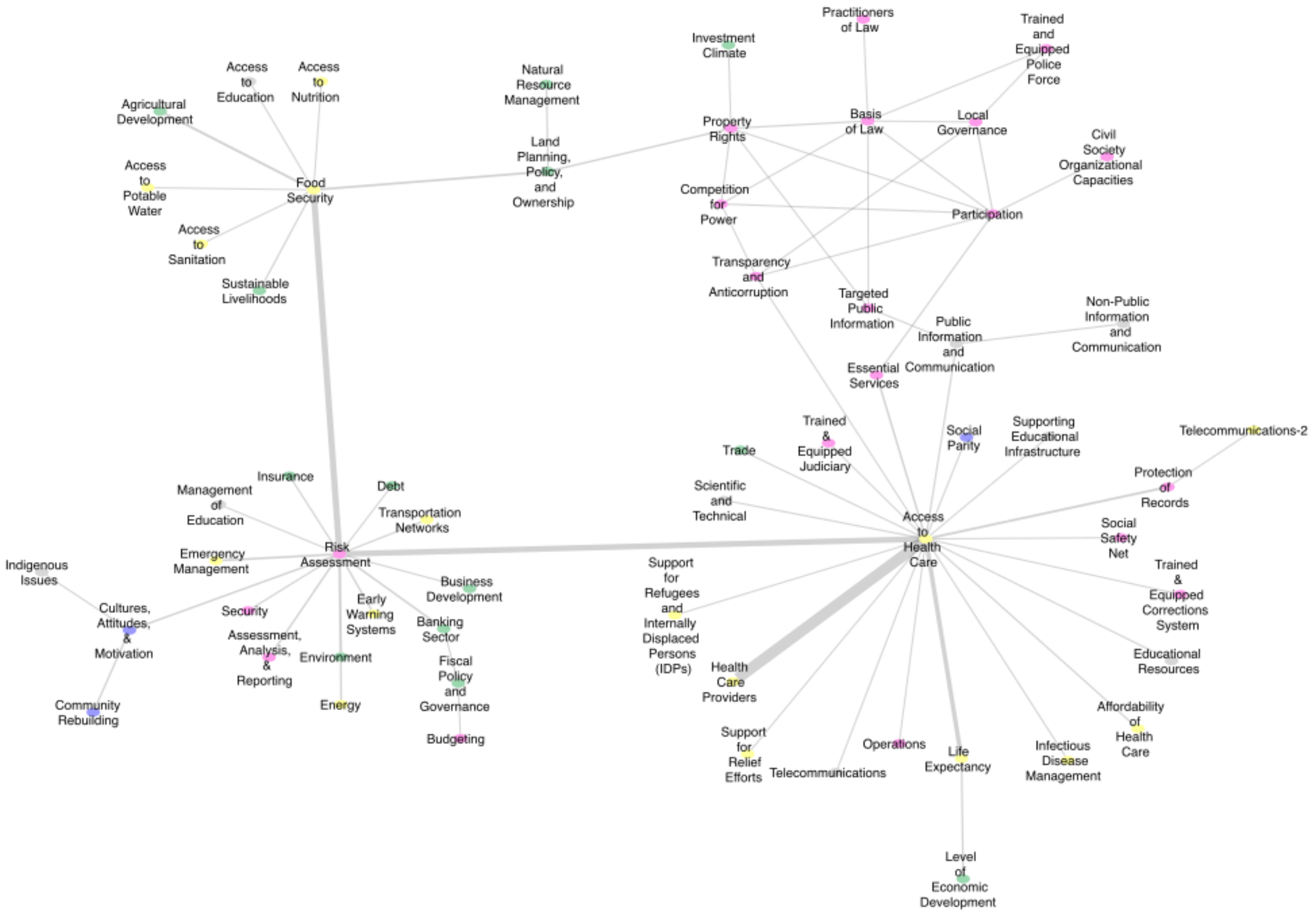

**Figure 4.** Complex Analysis for Social Adaptation: CASA Model. Node color represents sectors: green (Enable Resource Independence), yellow (Enhance Social Capacity), pink (Foster Good Governance), blue (Create Social Cohesion), and gray (Strengthen Knowledge Assets).

As shown in Fig. 4, the implemented methodology not only allowed the breaking of the basic structure of ARISE but also mixed elements from different sectors and pillars (see color of nodes).

# 3.- Results

### 3.1. Network Visualization and Analysis: Unveiling the Clusters and Connections of Social Resilience.

The outcomes of the methodology implemented are vividly depicted in network format in Fig. 4. This illustration reveals that the network is intricately organized into four principal clusters: *Health, Risk Management, Food Security, and Governance*. These clusters are interlinked, fostering the formation of a sophisticated social system dedicated to territorial resilience. It is imperative to underscore that the emergent structure of the network possesses a certain degree of universality. This universality is a direct consequence of the foundational methodology applied and is rooted in an extensive assimilation and application of global scientific knowledge spanning the last quarter-century. This approach underpins the genesis of connections between essential elements, illustrating the methodology's comprehensive and inclusive nature.

Additionally, Fig. 5 offers an insightful view of the network's higher-level connections (pillars) of the CASA model. These figures delineate the relationships organized by sectors and pillars, providing a deeper understanding of the network's overarching structure and the interplay of its constituent components.

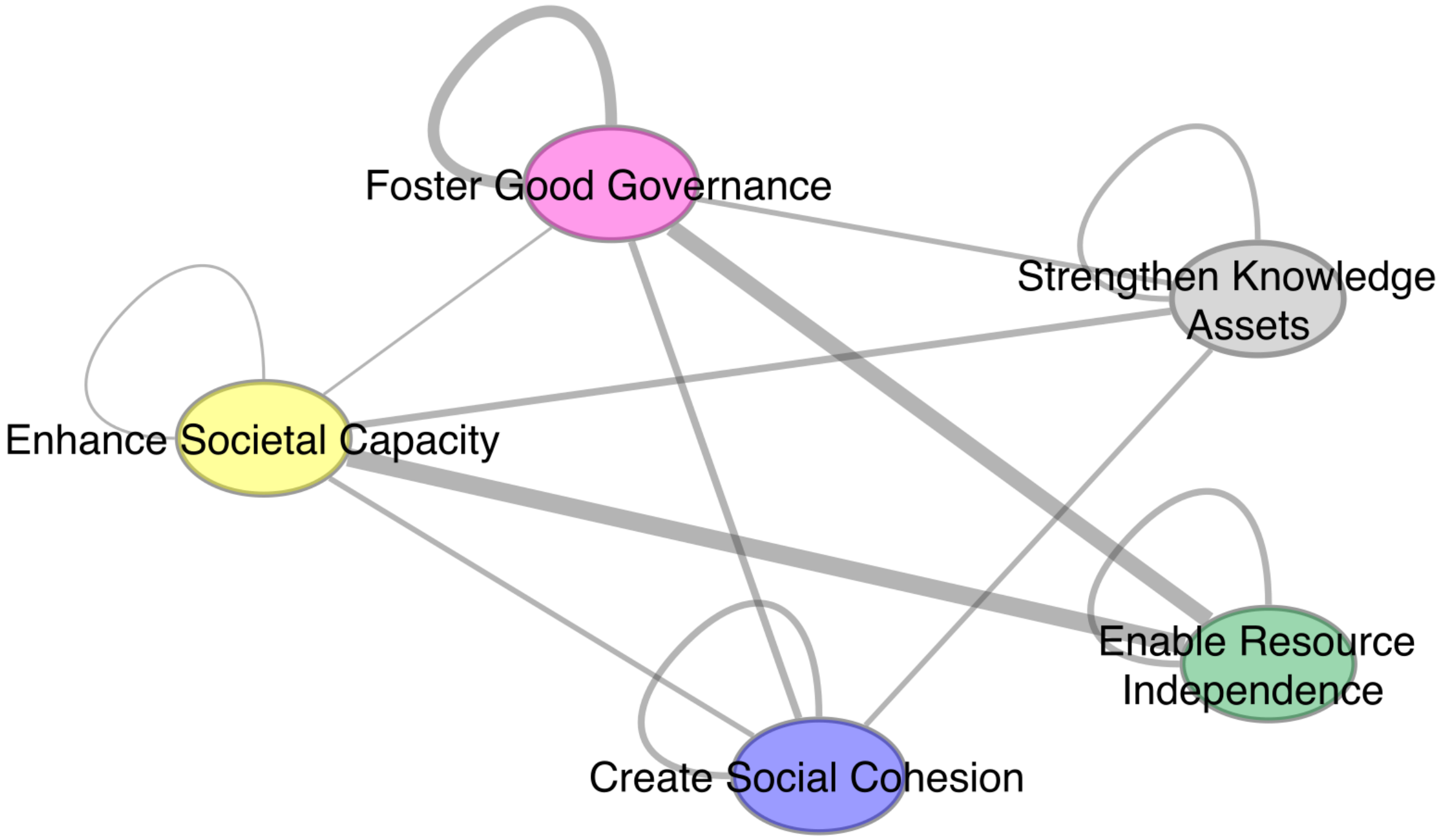

**Figure 5.** Network displaying the interaction at the upper level (pillars) of the CASA model.

### 3.2. Topological Resilience

CASA employs a topological framework to examine the social resilience of a territory. The positioning of components within this structure dictates their role in bolstering system resilience. Consequently, centrality metrics, including closeness and betweenness, are pivotal tools for pinpointing critical elements in the social resilience of a territory. These elements are characterized by their proximity to other components or function as intermediaries bridging different network sectors. Figure 6 illustrates the CASA model, highlighting nodes with the highest centrality regarding social resilience:

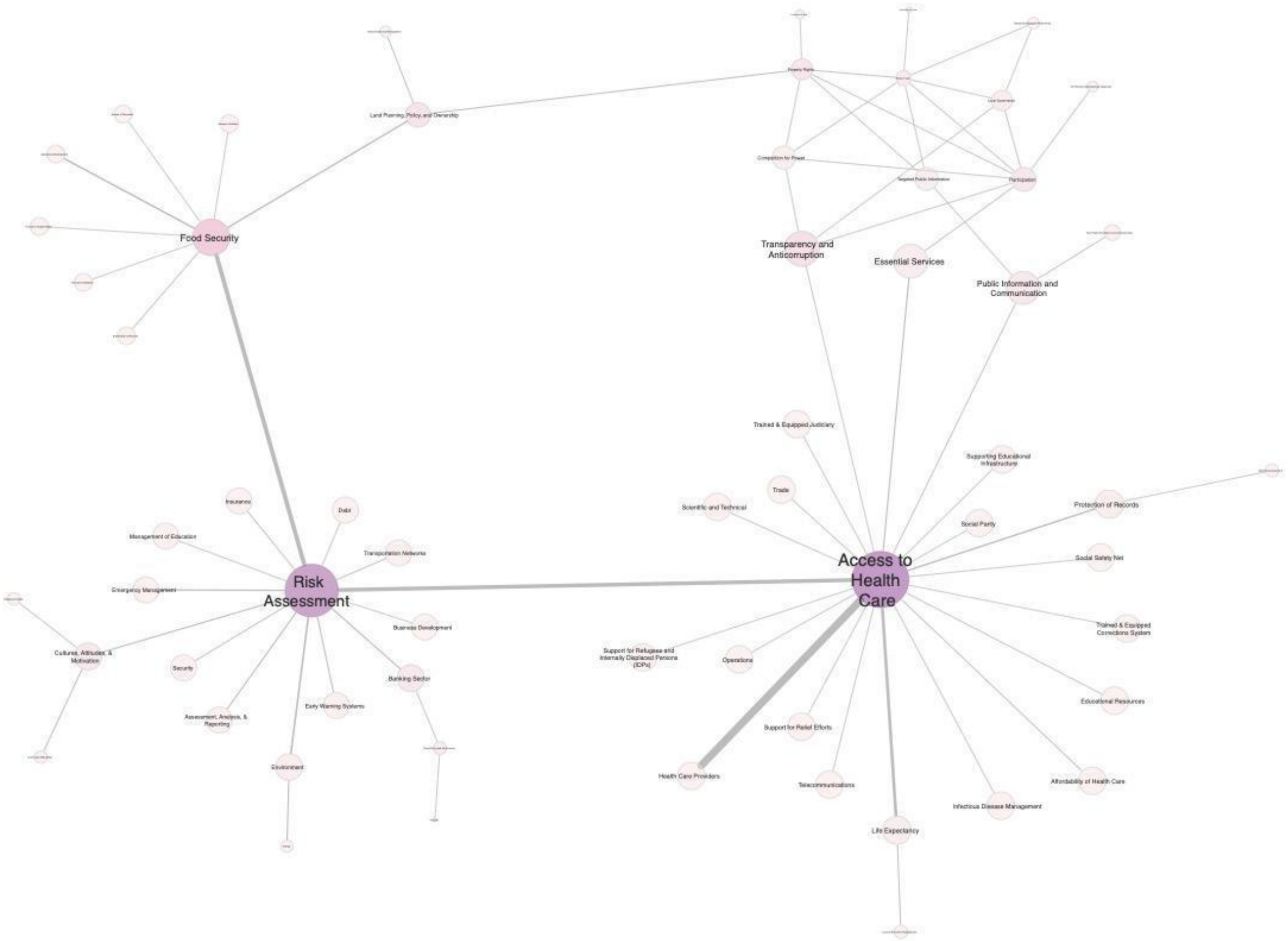

**Figure 6.** Component Centrality in the CASA Model: The size of the nodes and their labels indicate the relative closeness of each component to others within the model. The color gradient, transitioning from white to red, denotes the components' betweenness centrality, reflecting their intermediary role within the network's structure.

As can be observed, components of social resilience of a territory, such as *Access to Health Care* and *Risk Assessment*, along with others to a lesser extent, such as *Food Security*; *Transparency and Anticorruption*; *Essential Services* and *Public Information and Communication* are components that, for a given territory, must be pillars for the social resilience. It implies that in the face of a crisis, it must be there where the authorities and decision-makers must put in their greatest recovery efforts.

Considering the topological properties of the CASA model components, we introduce the concept of *Topological Resilience*, *TR*, which corresponds to the social resilience capacities installed in a territory for each of the components, $\rho_i$, and the location of these components, $c_i$, in the graph given by its proximity to other elements. Equation 1 shows the definition of a *TR* for a particular territory.

$$TR = \sum_N \left( \frac{N-1}{\sum_i d(i,j)} \right) \rho_i \quad (1)$$

The first part of the equation corresponds to the closeness centrality $c_i$ for component *i*, where *N* is the number of components of the CASA model, *d(i,j)* is the distance (length of the shortest path) between components *i* and *j*, and $\rho_i$ the capacities installed in component *i*. It is necessary to highlight that the topological resilience *TR* combines both the topology of the structure (*c*) and the capacities installed in that territory ($\rho$). This last component of the equations means that *TR* only makes sense for a particular territory since it depends on the data collected.

### 3.3. Dimensionality Reduction in the CASA Network: Identifying and Analyzing Super-Elements for Social Resilience

In the preceding sections, we have explored the development of a social resilience model, emphasizing the intricate systemic interrelationships among its various foundational elements. The derived structure enables the generation of emergent indicators at different levels, which are universally applicable due to the methodology employed, based on the scientific knowledge accumulated on the subject in the last twenty-five years. These indicators can range from individual, element-specific measures to neighborhood-level indicators (considering neighborhood definitions in network theory, not physical geography) or direct indicators based on predefined concepts such as sectors or pillars. We refer to these as "static indicators," as they remain unaffected by dynamic changes within the network, such as the impact of a catastrophic event on one or more nodes. Understanding the propagation of such an event is crucial, as its effects are likely to be more pronounced in nodes proximate to the initially impacted one. This observation underscores the potential for dynamic resilience analysis within the CASA model. That topic, while beyond the scope of this paper, is earmarked for future research.

Employing the *community detection algorithm* (Blodel et al., 2008; Lambiotte, 2009), we successfully simplified the complexity of the CASA Model into a more manageable form, comprising eight 'super-elements.' These super-elements represent distinct communities identified by the algorithm and are depicted in Figure 7.

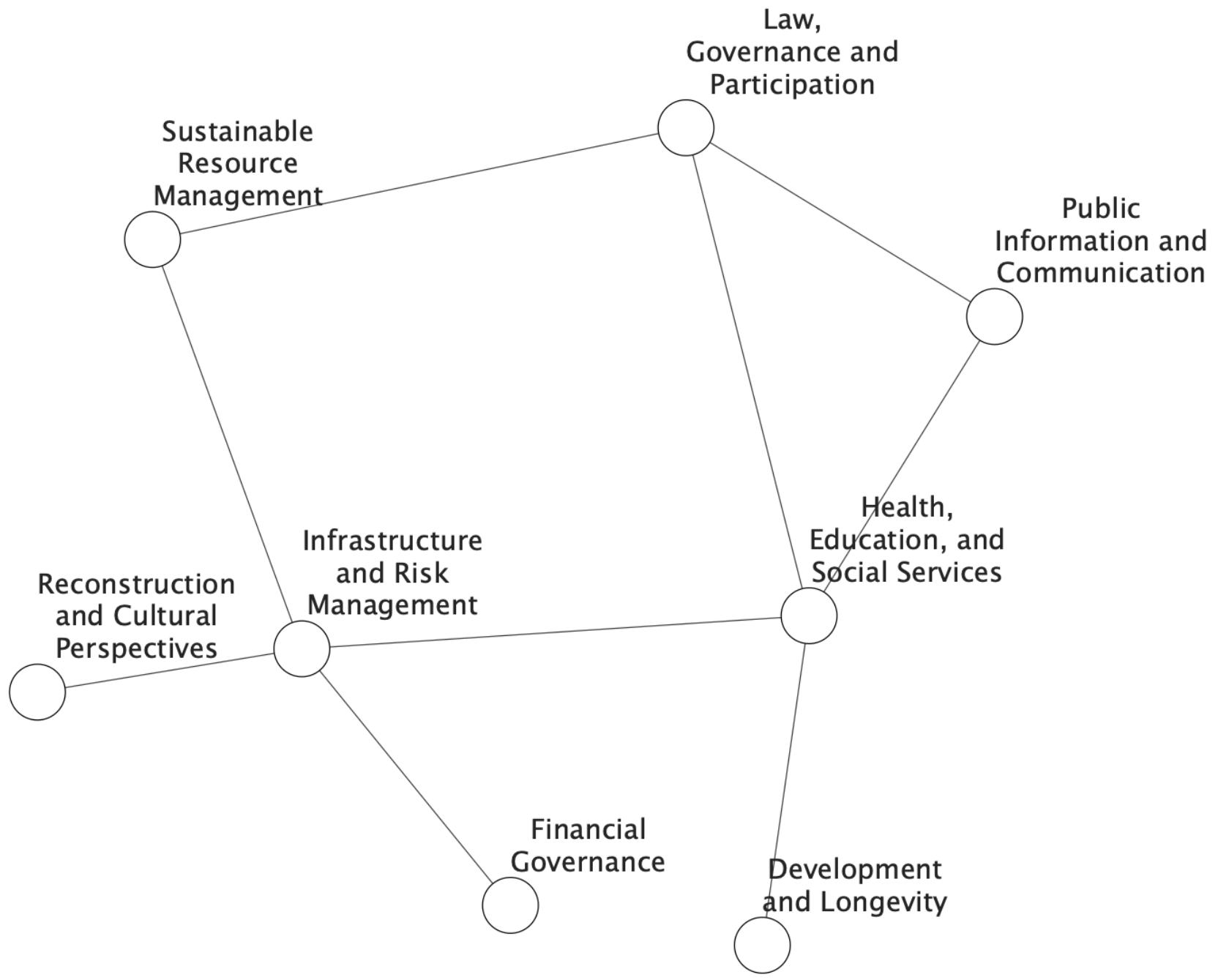

**Figure 7.** Super-elements in the CASA Model: Each node represents a concept based on the nature of the elements it encompasses.

These super-elements define emergent indicators and encapsulate and highlight the universal traits inherent in the interactions among the network's constituent elements. Table 2 shows the super-elements and the CASA components that belong to them.

Table 2. Super-elements of CASA Model.

| **CASA super-element** | **CASA elements** |
|---|---|
| Reconstruction and Cultural Perspectives | Community Rebuilding<br>Cultures, Attitudes, & Motivation<br>Indigenous Issues |
| Infrastructure and Risk Management | Assessment, Analysis, & Reporting<br>Business Development<br>Debt<br>Early Warning Systems<br>Energy<br>Emergency Management<br>Environment<br>Insurance<br>Management of Education<br>Risk Assessment<br>Security |
| | Transportation Networks |
| Sustainable Resource Management | Access to Education<br>Access to Potable Water<br>Access to Sanitation<br>Agricultural Development<br>Food Security<br>Land Planning, Policy, and Ownership<br>Natural Resource Management<br>Sustainable Livelihoods |
| Law, Governance, and Participation | Basis of Law<br>Civil Society Organizational Capacities<br>Competition for Power<br>Investment Climate<br>Local Governance<br>Participation by citizens<br>Practitioners of Law<br>Property Rights<br>Trained and Equipped Police Force<br>Transparency and Anticorruption |
| Public Information and Communication | Non-Public Information and Communication<br>Public Information and Communication<br>Targeted Public Information |
| Health, Education, and Social Services | Access to Health Care<br>Affordability of Health Care<br>Infectious Disease Management<br>Educational Resources<br>Essential Services<br>Health Care Providers<br>Operations for services delivery<br>Protection of Records<br>Scientific and Technical<br>Social Parity<br>Social Safety Net<br>Support for Refugees and Internally Displaced Persons (IDPs)<br>Support for Relief Efforts<br>Supporting Educational Infrastructure<br>Telecommunications<br>Telecommunications-2<br>Trained & Equipped Corrections System<br>Trade<br>Trained & Equipped Judiciary |
| Development and Longevity | Level of Economic Development<br>Life Expectancy |
| Financial Governance | Banking Sector |
| | Budgeting<br>Fiscal Policy and Governance |

This approach of dimensionality reduction offers a comprehensive view of the social resilience of a territory anchored primarily by four pivotal nodes:

- 'Infrastructure and Risk Management',
- 'Health, Education, and Social Services',
- 'Sustainable Resource Management', and
- 'Law, Governance, and Participation'.

The other four super-elements, while less central, maintain peripheral connections within this ring-structured framework.

Our subsequent analysis will focus on applying these topological results to the case of a community in a specific territory, which is affected by environmental degradation. This case study explores the viability of accruing territorial information through a bottom-up approach. Additionally, we will illustrate that aggregating data from larger territories is feasible and practical in the broader context of our study.

## 4. Applications of the CASA Model: A Sacrifice Zone

### 4.1. Topological Resilience: Impacts on Critical Sectors such as Health, Food Security, and Emergency Management

In this section, we present a static application of the model. Initially, we explore the quantitative application of the CASA model to a specific territory, involving actual territorial data to populate the various components of the model, denoted as $\rho$. Additionally, we discuss the methodology concerning central nodes ($c_i$) and their surrounding areas to deduce the territory's vulnerability.

In the town of Puchuncaví, an hour north of the Chilean city of Valparaiso, there is an industrial corridor that houses more than 15 companies within a 5 km strip along the coast, and each discharges an effluent into the air, the water, or the soil. Those discharges occur near ocean beaches, fresh-water rivers, farmlands, residential areas, and schools; each toxic effluent poses a risk to humans, animals, and plants (Bolados, 2016). The Puchuncaví commune is often characterized as a 'Sacrifice Zone' due to its enduring environmental stress, as highlighted by Hormazabal et al. (2019) and Vivanco (2022).

The social and environmental damage that occurs due to industrial pollution at Puchuncavi is a source of social instability. Mainly since 2018, the Puchuncaví area has had to face several episodes of health alerts, mainly due to poor air quality conditions due to high levels of sulfur, arsenic, and particulate matter in the area (Seguel et al., 2023), which can cause health problems that population manifest in symptoms such as fainting, nausea and vomiting. In fact, in August 2018, there were 1,370 emergency services associated with air quality. Due to the seriousness of what happened, the suspension of classes in 31 schools and 19 kindergartens was decreed in the commune (MINSAL, 2018). As a result of these types of events, which have been repeated since 2018, communities, health and education officials have formed a joint working group for *Mental Health and Education in Disaster Risk Management*, paying special attention to the repercussions that these events have not only on the physical health but also on the mental health of the people who inhabit that territory. Among the activities that have been deployed, there are emotional support activities for students and teachers, self-care workshops for teams, training in psychological first aid, talks and mental health training, among others.

These initiatives seek to promote the development of capacities for the protection of mental health and psychosocial support in emergency and disaster situations in all areas of risk management in the Educational Communities of Puchuncaví, including its neighboring town of Quintero. Among the objectives of these initiatives is to collaboratively build an intersectoral work plan to address these situations, encourage active citizen participation in the community in identifying risks, resources, and capacities, and generate articulation between schools and health teams. The conflicts around the Sacrifice Zones have been described as the inevitable consequence of an extractivist production and development model derived from neoliberal territorial planning (Bolados, 2016; Hormazábal et al., 2019).

During October and November 2023, data was amassed to operationalize the CASA Model for the Puchuncaví commune. The data was collected from both public information and through interviews with key actors and ethnographic visits to the place.

As a result of incorporating territorial data into the CASA topology, Fig. 8 illustrates the topological resilience *TR* of each component according to Eq. 1.

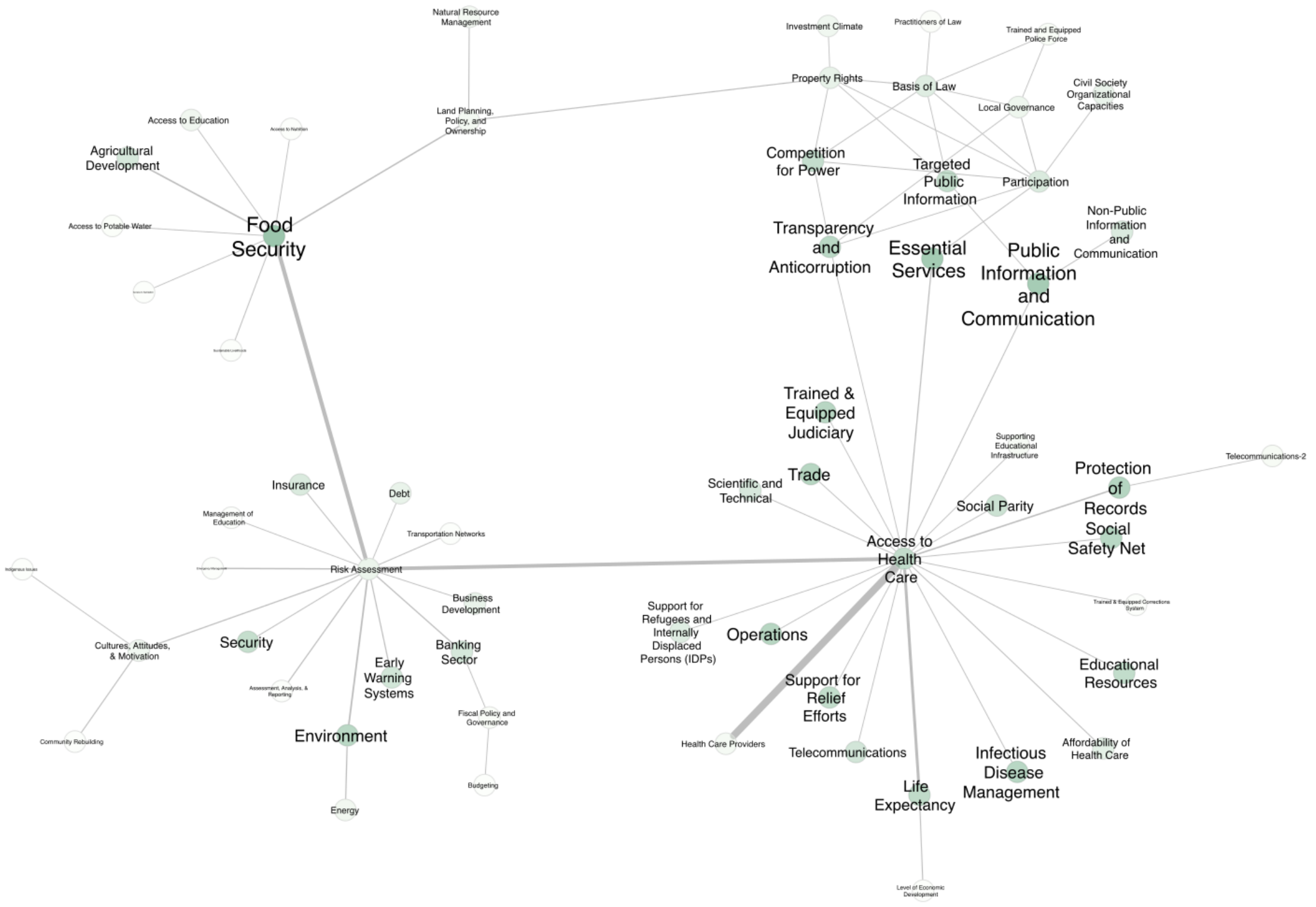

**Figure 8.** CASA Model for Puchuncavi. Vivid green color and larger labels represent nodes with greater installed capacities for Topological Resilience (*TR*).

The network analysis shows that the nodes vary in size, reflecting the installed capacities (collected values) in the territory for each CASA social resilience model component. In this case, the network representation shows, in a more vivid green color and with larger labels, those nodes in which Puchuncaví town has greater installed capacities for resilience, such as *Food Security, Essential Services, and Public Information and Communication*.

Conversely, in Fig. 9 with red colors and larger labels, low values indicate vulnerabilities in the system in aspects such as *Indigenous Issues, Access to Nutrition, Sustainable Livelihoods, Access to Sanitation, and Emergency Management*.

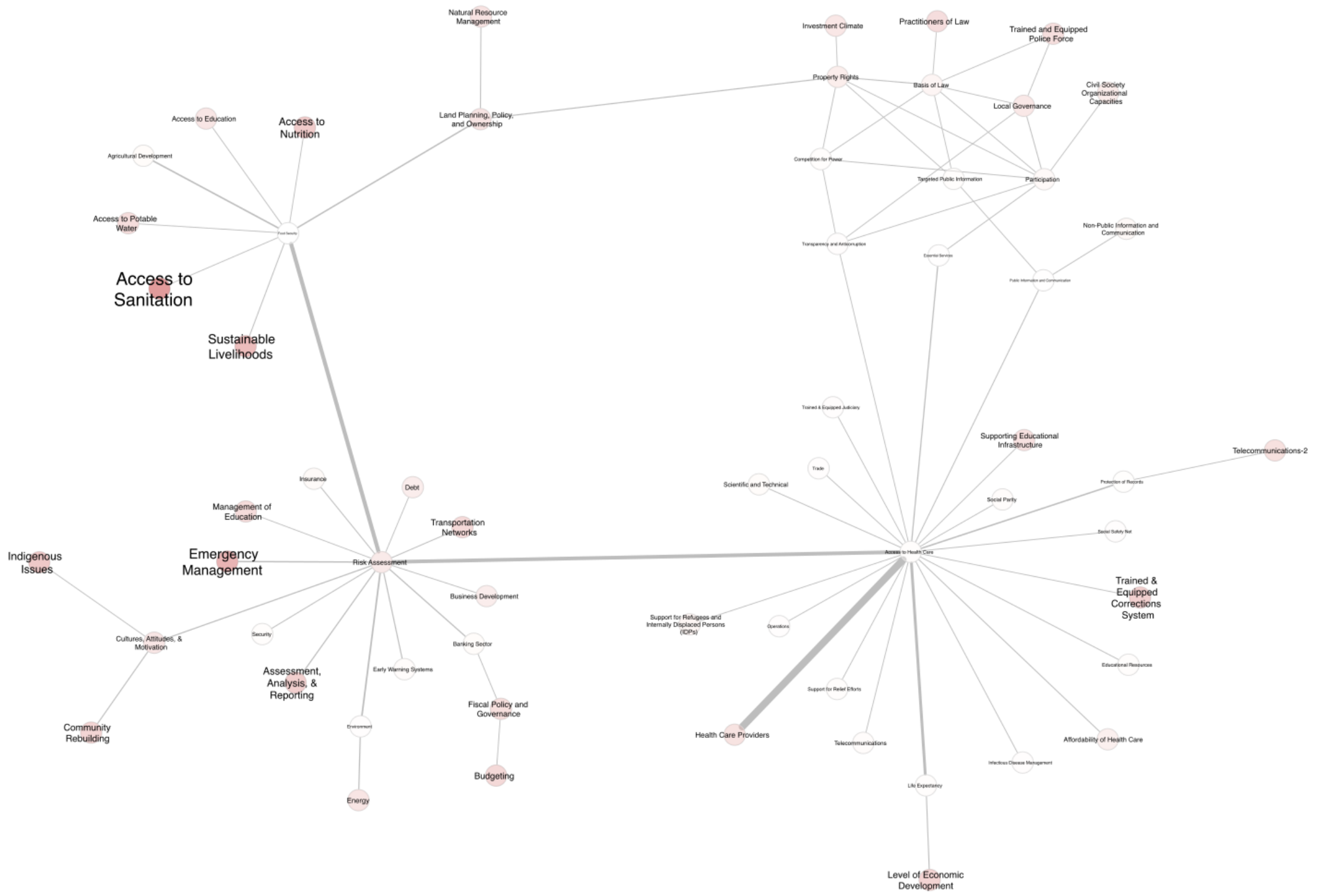

**Figure 9.** CASA Model for Puchuncavi. The vivid red color and larger labels represent nodes with vulnerabilities in Topological Resilience (*TR*).

To simplify this representation, we calculated the mean value of the elements (nodes) comprising each super-element (Fig. 10).

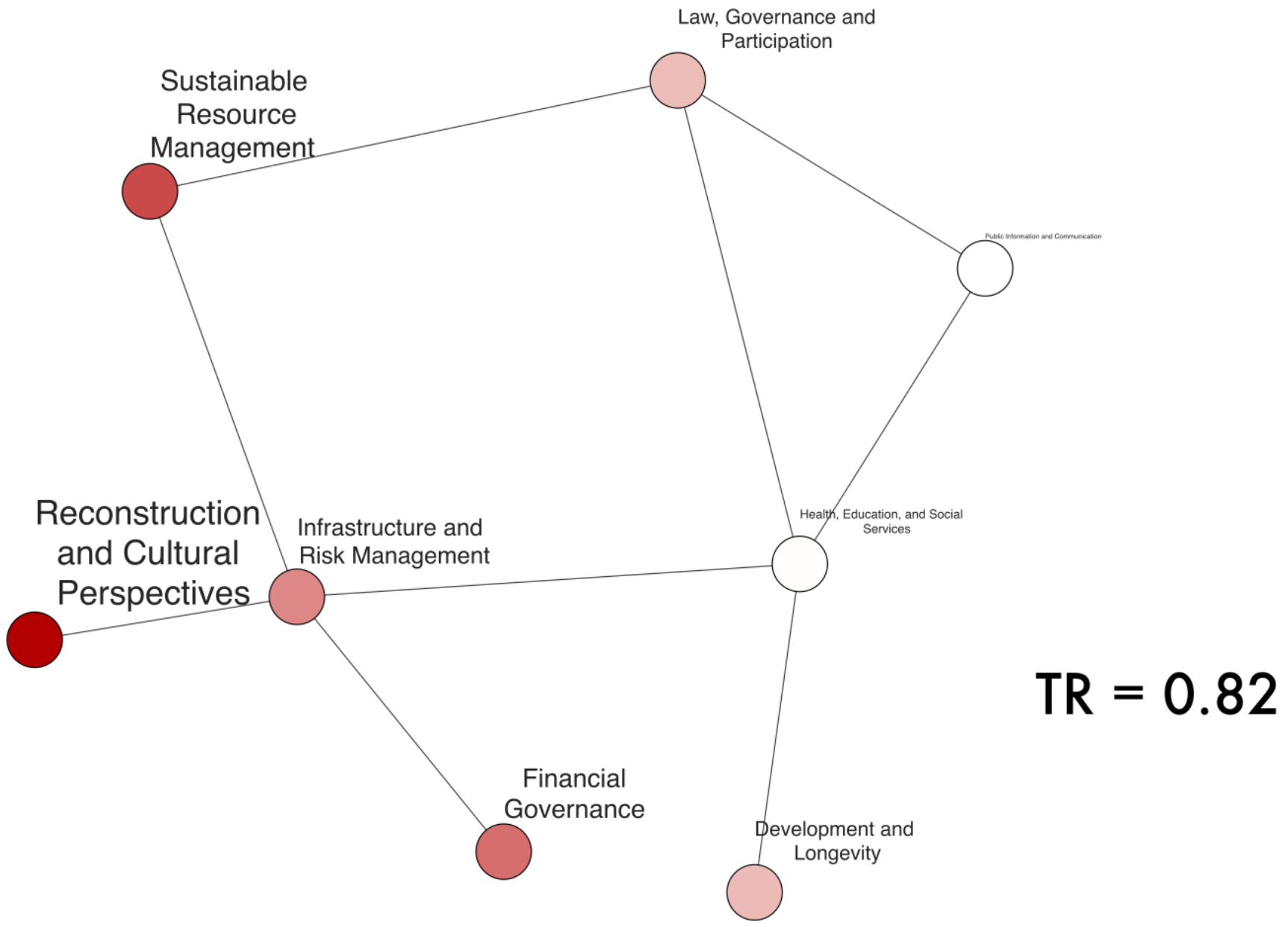

**Figure 10.** Resilience structure for Puchuncavi as seen in CASA Model super-elements. The vivid red color and larger labels represent super-elements with vulnerabilities in Topological Resilience (*TR*).

It can be discerned that the town of Puchuncaví possesses established capacities to mitigate disturbances within the realms of *Health*, *Education*, and *Social Services*, as well as *Public Information and Communication* and *Law*, *Governance*, *and Participation*. Conversely, the territory exhibits vulnerabilities in *Reconstruction and Cultural Perspectives*, *Sustainable Resource Management*, *Financial Governance*, and *Infrastructure and Risk Management*. The extant capacities within the territory endow Puchuncaví with a Territorial Resilience value of *TR*=0.82 where the maximum resilience is 1.0, signifying that Puchuncavi is 18% more susceptible to adverse events than a territory endowed with the maximal level of topological resilience capacities.

### 4.2. Evaluation of Critical Events: An Example Linked to the Climate Crisis

A second application pertains to one of the primary aims for which the CASA model was conceived: enhancing the social resilience of territories in the face of the climate crisis. Employing the *Scientifically Informed Connectivity* approach, we analyzed a comprehensive corpus of scientific literature from the Web of Science from 2008 to 2023.

This analysis concentrated on the co-occurrence of precise keywords that delineate the elements incorporated within the CASA model alongside a set of keywords indicative of pivotal events within climate phenomena. For this co-occurrence analysis, the scope was specifically limited to instances situated within the context of the climate crisis (i.e., co-occurrences were considered relevant when they transpired in conjunction with the terms "Global Warming" and "Climate Change"). Figure 11 exemplifies the influence of the CASA model on the climate phenomenon of "drought.”

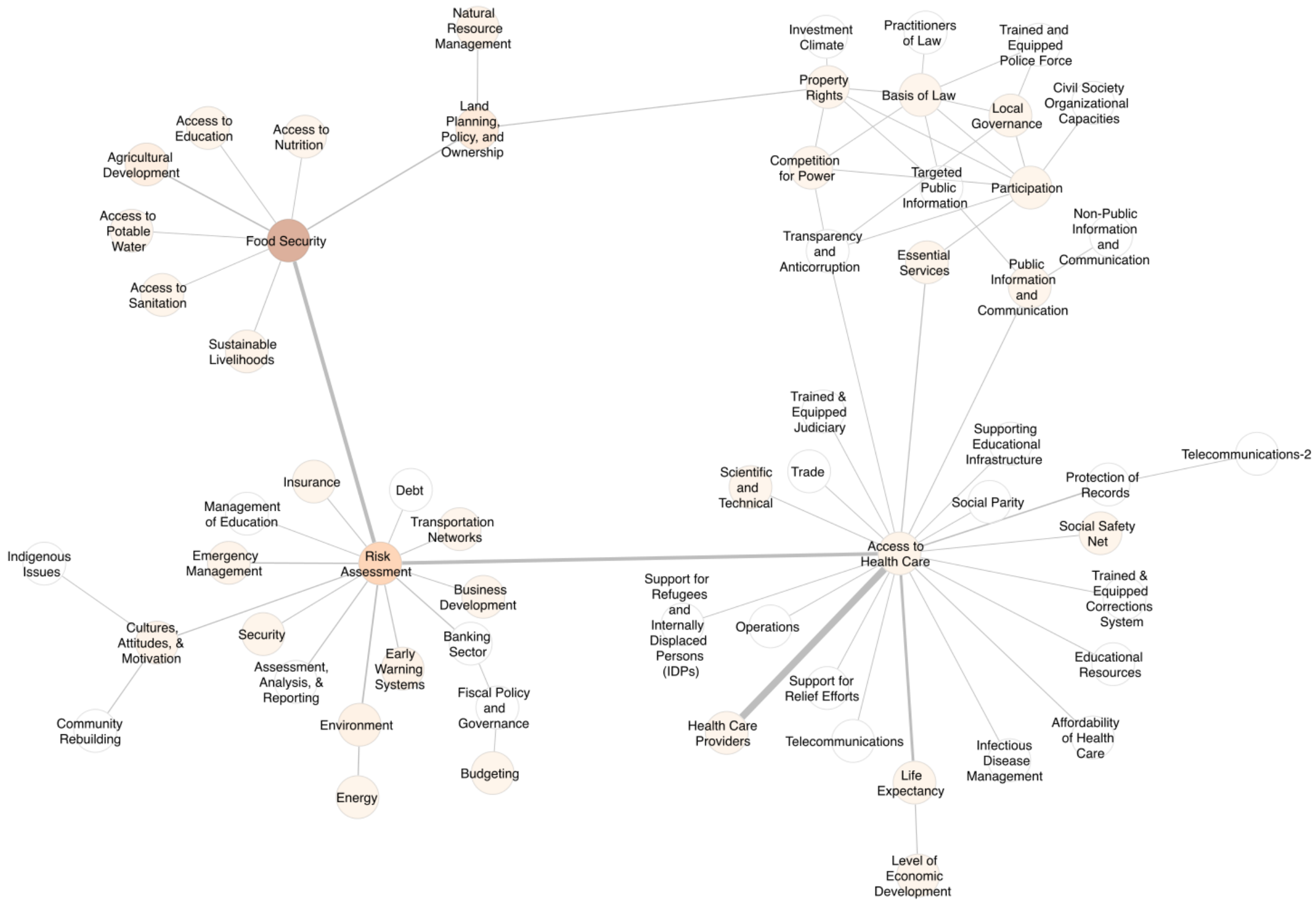

**Figure 11.** The CASA model illustrates the consequences of drought. The node color scale (white -> dark orange) represents the impact of climatic phenomena according to the *Scientifically Informed Connectivity* method discussed above.

As depicted, the *Food Security and Risk Assessment* components exhibit the most substantial correlation with the drought phenomenon. Nevertheless, it is evident that other system components, such as access to nutrition, education, potable water, and sanitation, are also impacted. Additional components like *Emergency Management* or *Access to Health* are directly associated with the phenomenon, which is currently experiencing frequency, duration, and intensity changes across certain territories.

At this juncture, by cross-referencing the topological resilience data for Puchuncaví (Fig. 8) with the environmental impact (Fig. 11), it is discernible that a phenomenon of this nature, which is already severely impacting the locality (Biblioteca del Congreso Nacional de Chile, 2018), has a significant direct effect on the foundational pillars of territorial resilience in the CASA model (Fig. 6), such as food security, access to health, local governance, and risk management. Conversely, in the context of Puchuncaví, this directly impinges upon one of its most defining socioeconomic components: agriculture, with its intrinsic link to food security.

## 5.- Conclusions

This research aimed to derive a rich and responsive social resilience model that considers the dynamic organizational complexity of human societies and their territories. For this, the *Analysis of Resilience for Integrated System Effectiveness* framework (ARISE Model) was used as a baseline design, which was transformed through the use of global scientific knowledge methodology and tools from the theory of complex systems into a topological resilience model called CASA (Complex Analysis for Social Adaptation).

Regarding the study of the base model and the development process of the CASA model, we conclude that the previous framework brought together social, political, and economic dimensions, among others, to quantify a territory's social resilience. However, such a quantification remained indebted to the systemic approach since it ignored the relationships between the components of the territorial system.

In that sense, the CASA Model introduces complexity and provides insight into a resilient social system based on global descriptions in the relevant scientific literature and ground-based assessments in high-threat areas. The CASA Model could represent a universal resilience architecture of social systems, particularly in its components and relationships. These relationships allow the introduction of concepts such as a *Topological Resilience Indicator*, where the location of the components of territorial resilience in such a structure denotes levels of importance. In this case, *Access to Health Care*, *Risk Assessment* and *Food Security* seem to be the key components defining a territory's capabilities to absorb and adapt to crises. They are also key to response efforts when a crisis affects the territory.

In related research, we have found that using meticulous prompts within appropriate AI tools (e.g. Anthropic's “Claude”, an LLM developed on a constitutional framework) can be valuable in identifying new indicators for dimensionality reduction. In this sense, the analysis of communities allowed us to detect eight super-elements that reduce the dimensionality of the model and allow us to have a clearer vision of the resilience capacities of a territory. This simplified look at 'Infrastructure and Risk Management'; 'Health, Education, and Social Services'; 'Sustainable Resource Management', and 'Law, Governance, and Participation' are the key components of territorial resilience.

The results obtained from the use of the CASA model for a territory like Sacrifice Zone seem to be promising in several aspects: for describing the topological resilience of the territory, for evaluating the impacts on it of phenomena of any type, such as environmental ones derived from climate change; and to support community organization actions to confront crises, many of them related to human health, such as the "Working Groups for Mental Health and Education in Disaster Risk Management". In the case of the description of the topological resilience of the town under study, the results show its character as a Zone of Sacrifice. Given this scenario, it is not surprising that aspects such as *Public Information and Communication*, *Social Security Net,* and *Access to Health Care* are strong given the need to communicate the constant risk in which their population finds themselves, as well as the ability to provide access to health on time, all within a predominantly agricultural territory where food security is threatened by chemical contamination.

These strengths contrast with those weaknesses in aspects such as *Emergency Management*, a clear reflection of the territory's inability to improve its response to the constant poisoning experienced by its population. On the other hand, the territory seems to be weak in other aspects, such as *Sustainable Livelihoods, and in* access to nutrition, sanitation, and potable water, possibly in the latter case due to the drought that has affected the area for decades.

This Sacrifice Zone scenario translates into higher-scale vulnerabilities, denoted by the super-elements of the CASA model. In fact, *Reconstruction and Cultural Perspectives*, *Sustainable Resource Management*, and Infrastructure and Risk Management appear as weak super-elements, perhaps from a long history of social unrest due to persistent environmental damage, drought, and soil contamination that threaten their food security and livelihoods.

The simulation of the climatic impact of the drought on the structure of CASA and the town of Puchuncaví shows the pressure on the pillars of topological resilience of this CASA framework. In the case of Puchuncaví, the simulation reveals the pressure that this phenomenon exerts on one of its identity pillars, such as agriculture, and the chain of effects that can ensue. On the other hand, the simulation shows the fragility of the town as Risk Management is revealed as a weak component in Puchuncaví while at the same time, the town is profoundly affected by drought..

This work opens several lines of research. We intend to improve the proposed model by refining the most revealing elements, increasing or eliminating relationships between them, and reviewing the data that feeds the territorial model.

In parallel, a critical theme for future research is the analysis of production and consumption systems, which illuminates the material and energetic flows foundational to societal operations, carrying profound implications for environmental sustainability and societal well-being (Geels, McMeekin, Mylan, & Southerton, 2015). The shift towards sustainable production and consumption practices is acknowledged as essential in mitigating environmental degradation while enhancing economic and social resilience (Cohen, Brown, & Vergragt, 2010).

The interconnection and dynamic interaction between social resilience models and production and consumption systems become apparent within the sustainable livelihoods framework, which amalgamates the resilience of social structures with sustainable economic activities. This approach evaluates how economic endeavors, entrenched in resilient social frameworks, bolster the capability of individuals and communities to confront and adapt to environmental and societal shifts (Scoones, 1998). Such a viewpoint is in harmony with the circular economy's principles, advocating for a systemic alteration in production and consumption modalities towards resource efficiency and waste reduction, thus contributing simultaneously to environmental sustainability and social resilience (Stahel, 2016).

Moreover, adaptive governance is paramount in connecting these areas. Adaptive governance models underline the necessity for flexible, participatory, and multi-tiered governance mechanisms adept at addressing complex socio-environmental predicaments through fostering stakeholder collaboration and amalgamating diverse knowledge systems (Folke et al., 2005). These governance frameworks are vital in steering the transition to sustainable production and consumption systems, as they facilitate the collaborative crafting of policies and practices that resonate with the socio-ecological nuances of diverse communities (Chaffin, Gosnell, & Cosens, 2014).

Finally, we believe that studying damage propagation models within the CASA framework is the next logical step. We have already assigned an internal research group to this analysis.

Once refined, we intend to share the CASA framework as an interactive tool open to the public. This should enable global communities to explore their resilience as they see fit, adjusting budgets, policies, and urban planning decisions accordingly.

## List of abbreviations

ARISE: Analysis of Resiliency for Integrated System Effectiveness; CAS: Complex Adaptive Systems; CASA: Complex Analysis for Socio-environmental Adaptation.

## Declarations

Availability of data and materials: The datasets used and/or analysed during the current study are available from the corresponding author on request.

Competing interests: The authors declare that they have no competing interests.

Funding: This work was supported by the Office of Naval Research (ONR) under Grant #N00014-22-1-2613 and FONDECYT 1211323.

Authors' contributions: Conceptualization, M.F. and J.P.C.; methodology, M.F. and J.P.C.; software J.P.C. and G.O.; investigation, M.F., J.P.C., C.U., G.V., E.R. and S.S.; data curation, C.U. and G.O.; writing—original draft preparation, M.F. and J.P.C.; writing—review and editing, M.F., C.U., E.R. and J.P.C. All authors have read and agreed to the published version of the manuscript.

Acknowledgements: The authors sincerely appreciate our stimulating discussions with Dr. John Hummel, the architect of the Analysis of Resilience for Integrated System Effectiveness model, who kindly encouraged us to modify it.